\documentclass{elsart}

\usepackage{natbib}

\usepackage{epsfig}

\usepackage{amssymb}

\begin{document}

\begin{frontmatter}

% Title, authors and addresses

% use the thanksref command within \title, \author or \address for footnotes;
% use the corauthref command within \author for corresponding author footnotes;
% use the ead command for the email address,
% and the form \ead[url] for the home page:
% \title{Title\thanksref{label1}}
% \thanks[label1]{}
% \author{Name\corauthref{cor1}\thanksref{label2}}
% \ead{email address}
% \ead[url]{home page}
% \thanks[label2]{}
% \corauth[cor1]{}
% \address{Address\thanksref{label3}}
% \thanks[label3]{}

\title{Models for the IR and submm SEDs of normal, starburst and active galaxies}
\author{Michael Rowan-Robinson} 
\address{Astrophysics Group, Imperial College London, Blackett Laboratory,
Prince Consort Road, London SW7 2BZ}

\begin{abstract}
Models for the seds of infrared galaxies are reviewed, with particular application to hyperluminous
infrared galaxies.  

\end{abstract}

\end{frontmatter}

%\large

\section{Ingredients for model SEDs}
The key ingredients for models for the spectral energy distributions (seds)
of infrared galaxies are 
{\bf(1) A model for interstellar dust grains.}  
Detailed models have been provided by Mathis et al (1977), Draine and Lee (1984), 
Rowan-Robinson (1986, 1992), 
Desert et al (1990), Siebenmorgen and Krugel (1992), Dwek (1998).
{\bf(2) An assumed density distribution for the dust.}
Detailed physical density profiles have been developed by Yorke (1977) for HII regions, by 
Efstathiou and Rowan-Robinson (1991) for protostars and by Eftathiou et al (2000) for starbursts.  
However most authors have followed the approach of
Rowan-Robinson (1980), who explored models with N(r) $\propto r^{-\beta}$.
Silva and Danese (1998) have investigated exponential density profiles for the cirrus component.
{\bf(3) The geometry of the dust.}
The most common assumption is spherical symmetry.  Axi-symmetric models have been investigated by
Efstathiou and Rowan-Robinson (1990, 1991, 1995),
Pier and Krolik (1992), Granato et al (1994, 1997), Silva et al (1998).  
Rowan-Robinson (1995) gave an approach to modelling emission from
an ensemble of discrete compact clouds. 
{\bf(4) A radiative transfer code}
The first accurate spherically symmetric radiative transfer code was by Rowan-Robinson (1980)
and the first accurate axisymmetric code by Efstathiou et al (1990).  Pier and Krolik (1992),
Krugel and Siebenmorgen (1994), Granato et al (1997) and Silva et al (1998) have also
developed axisymmetric codes.
{\bf(5)Source components}
Typically, seds can be understood as a superposition of several distinct source components.
Rowan-Robinson and Crawford (1989) were able to fit the IRAS colours and spectral
energy distributions of galaxies detected in all 4 IRAS bands with a mixture of 3 components,
emission from interstellar dust ('cirrus'), a starburst and an AGN dust torus.  Recently
Xu et al (1998) have shown that the same 3-component approach can be used to fit
the ISO-SWS spectra of a large sample of galaxies.  To accomodate the Condon et al (1991)
and Rowan-Robinson and Efstathiou (1993) evidence for higher optical depth starbursts, 
Ferrigno et al (2000, in prep.) have extended the Rowan-Robinson and Crawford (1989) analysis
to include a fourth component, an Arp220-like, high optical depth starburst, for
galaxies with log $L_{60} >$ 12.  Efstathiou et al (2000) have given improved
radiative transfer models for starbursts as a function of the age of the starburst,
for a range of initial dust optical depths, and similar analysis has been carried out by
Silva et al (1998) and Siebenmorgen (1999, 2000 - these proceedings).

Sanders et al (1988) proposed, on the basis of spectroscopic arguments for a sample of 10 objects,
that all ultraluminous infrared galaxies contain an AGN and that the far infrared emission is
powered by this.  Sanders et al (1989) proposed a specific model, in the context of
a discussion of the infrared emission from PG quasars, that 
the far infrared emission comes from the outer parts of
a warped disk surrounding the AGN.  This is a difficult hypothesis to disprove,
because if an arbitrary density distribution of dust is allowed at large distances from the
AGN, then any far infrared spectral energy distribution could in fact be generated.
Rowan-Robinson (2000a) discussed whether the AGN dust torus model of Rowan-Robinson (1995) 
can be extended naturally to explain the far infrared and submillimetre emission
from hyperluminous infrared galaxies, but concluded that in many cases this does
not give a satisfactory explanation.  
Rigopoulou et al (1996) observed a sample of ultraluminous infrared galaxies from
the IRAS 5 Jy sample at submillimetre wavelengths, with the JCMT, and at X-ray wavelengths, 
with ROSAT.  They found that most of the far infrared and submillimetre spectra were fitted well
with the starburst model of Rowan-Robinson and Efstathiou (1993).  The ratio of bolometric
luminosities at 1 keV and 60 $\mu$m lie in the range $10^{-5} - 10^{-4}$ and are
consistent with a starburst interpretation of the X-ray emission in almost all cases.
Even more conclusively, Genzel et al (1998) have used ISO-SWS spectroscopy to show that 
the majority of ultraluminous ir galaxies are powered by a starburst rather than an AGN.

\section{Applications to quasars and hyperluminous infrared galaxies}
Quasars and Seyfert galaxies, on the other hand, tend to show a characteristic mid 
infrared continuum, broadly flat 
in $\nu S_{\nu}$ from $3-30 \mu$m.  This component was modelled by 
Rowan-Robinson and Crawford (1989) as dust in the 
narrow-line region of the AGN with a density distribution n(r) $\alpha$  $r^{-1}$ .  
More realistic models of this 
component based on a toroidal geometry are given by Pier and Krolik 
(1992), Granato and Danese (1994), Efstathiou and 
Rowan-Robinson (1995).  Rowan-Robinson 
(1995) suggested that most quasars contain both (far ir) starbursts and (mid ir) 
components due to (toroidal) dust 
in the narrow line region.

One of the major discoveries of the IRAS mission was the existence of ultraluminous 
infrared galaxies, 
galaxies with $L_{fir} > 10^{12} h_{50}^{-2} L_{\odot} (h_{50} = H_{o}/50)$.  
The peculiar Seyfert 2 galaxy Arp 220 was recognised as having an exceptional far infrared 
luminosity early in the mission (Soifer et al 1984).  

The conversion from far infrared luminosity to star formation rate has been discussed 
by many authors (eg Scoville and Young 1983, Rowan-Robinson et al 1997).
Rowan-Robinson (2000b, in prep.) has given an updated estimate of how the star-formation
rate can be derived from the far infrared luminosity, finding

$\dot{M}_{*,all} /[L_{60}/L_{\odot}]$ = 2.2 $\phi/\epsilon$ x$10^{-10}$ 

where $\phi$ takes account of the uncertainty in the IMF (= 1, for a standard
Salpeter function) and $\epsilon$ is the fraction of uv light absorbed by dust, estimated
to by 2$/$3 for starburst galaxies (Calzetti 1998).
We see that the star-formation rates in
ultraluminous galaxies are $ > 10^{2} M_{\odot} yr^{-1}$.  However the time-scale
of luminous starbursts may be in the range $10^7 - 10^8$ yrs (Goldader et al 1997), so the
total mass of stars formed in the episode may typically be only 10$\%$ of the mass of a galaxy.

Here I discuss an even more extreme class of infrared galaxy, hyperluminous infrared
galaxies, which I define to be those with rest-frame infrared (1-1000 $\mu$m) luminosities, $L_{bol,ir}$,
in excess of $10^{13.22} h_{50}^{-2} L_{\odot}$ (=$10^{13.0} h_{65}^{-2} L_{\odot}$).  
For a galaxy with an M82-like starburst spectrum this corresponds to  
$L_{60} \geq 10^{13} h_{50}^{-2} L_{\odot}$, since the bolometric correction at 60 $\mu$m is 1.63.
  While the emission at
 rest-frame wavelengths 3-30 $\mu$m in these galaxies is often due to an AGN dust torus (see below),
 I argue that their emission at rest-frame wavelengths
$\geq 50 \mu$m is primarily due to extreme starbursts, implying star formation rates
in excess of 1000 $M_o/yr$.  These then are excellent candidates for being primeval galaxies, galaxies
undergoing a very major episode of star formation.  More details are given in Rowan-Robinson (2000a).

For a small number of these galaxies we have reasonably detailed continuum spectra from 
radio to uv wavelengths.  
Figures 1-6 show the infrared continua of some of these hyperluminous galaxies, with fits using radiative 
transfer models (specifically the standard M82-like starburst model and
an Arp220-like high optical depth starburst model from Efstathiou et al 
(2000) and the standard AGN dust torus model of Rowan-Robinson (1995).  

I now discuss the individual objects in turn:

$\bf{F10214+4724}$

The continuum emission from F10214 was the subject of a detailed discussion by Rowan-Robinson 
et al (1993).  
Green and Rowan-Robinson (1995) have discussed starburst and AGN dust tori models for F10214.  
Fig 1 shows M82-like and Arp 220-like starburst
models fitted to the sumbillimetre data for this galaxy.  The former gives a
good fit to the latest data.  The 60 $\mu$m flux requires an AGN dust torus
component.  To accomodate the upper limits at 10 and 20 $\mu$m, it is necessary
to modify the Rowan-Robinson (1995) AGN dust torus model so that the maximum temperature
of the dust is 1000 K rather than 1600 K.  I have also shown the effect of allowing the
dust torus to extend a further factor 3.5 in radius.  This still does not account for the amplitude
of the submm emission.  The implied extent of the narrow-line region for this extended AGN
dust torus model, which we 
use for several other objects, would be 326 $(L_{bol}/10^{13} L_{\odot})^{1/2}$ pc 
consistent with 60-600 $(L_{bol}/10^{13} L_{\odot})^{1/2}$ pc quoted by Netzer (1990).
  Evidence for a strong starburst contribution to the ir emission from F10214
is given by Kroker et al (1996) and is supported by the high gas mass detected via
CO lines.  Granato et al (1996) attempt to model the whole sed of F10214
with an AGN dust torus model, but still do not appear to be able to account for the 60 $\mu$m
emission.

$\bf{SMMJ02399-0136}$

A starburst model fits the submm data well and the ISO detection at 15 $\mu$m
gives a very severe constraint on any AGN dust torus component.  The starburst 
interpretation of the submm emission is supported by the gas mass estimated from CO detections
(Frayer et al 1998).

$\bf{F08279+5255}$

An M82-like starburst is a good fit to the submm data and an AGN dust torus model is a good
fit to the 12-100 $\mu$m data.  The high gas mass detected
via CO lines (Downes et al 1999) supports a starburst interpretation, though the ratio of 
$L_{sb}/M_{gas}$ is on the high side.  The submm data can also
be modelled by an extension of the outer radius of the AGN dust torus.

$\bf{H1413+117}$

The submm data is well fitted by an M82-like starburst and the gas mass implied by the 
CO detections (Barvainis et al 1994) supports this interpretation.  The extended AGN dust torus model
discussed above does not account for the submm emission.  However Granato et al
(1996) model the whole sed of H1413 in terms of an AGN dust torus model.

$\bf{15307+325}$

A starburst model gives a natural explanation for the 60-180 $\mu$m excess compared
to the AGN dust torus model required for the 6.7 and 14.3 $\mu$m emission (Verma et al 
2000), but the non-detection of CO poses a problem for a starburst intepretation.

$\bf{PG1634+706}$

The IRAS 12-100 $\mu$m data and the ISO 150-200 $\mu$m data (Haas et al 1998) are well-fitted 
by the extended AGN dust torus model
and an upper limit can be placed on any starburst component.  The non-detection of CO is
consistent with this upper limit.

 In Fig 7 we show the far infrared 
luminosity derived for an assumed starburst component, versus 
redshift, for hyperluminous galaxies, with lines indicating observational 
constraints at 60, 800 and 1250 $\mu$m.  Three of the sources with 
(uncorrected) total bolometric luminosities 
above $10^{14} h_{50}^{-2} L_{o}$ are strongly gravitationally 
lensed.  IRAS F10214+4724
was found to be lensed with a magnification which ranges from 100 at optical wavelengths
to 10 at far infrared wavelengths (Eisenhardt et al 1996, Green and Rowan-Robinson 1996).
The 'clover-leaf' lensed system H1413+117 has been found to have a magnification of
10 (Yun et al 1997).  Downes et al (1999) report a magnification of 14 for F08279+5255.
Also, Ivison et al (2000)  estimate a magnification of 2.5 for SMMJ02399-0136 and Frayer et al (1999)
quote a magnification of 2.75 $\pm$0.25 for SMMJ14011+0252.
These magnifications have to be corrected for in estimating luminosities (and dust and
gas masses) and these corrections are indicated in Fig 7. 

% Have here your figure (s) included as "file" in the \epsfig statement
% Have one \begin ...\end per figure
% Of course this figure can be inserted somewhere in the text....

\begin{figure}
\epsfig{file=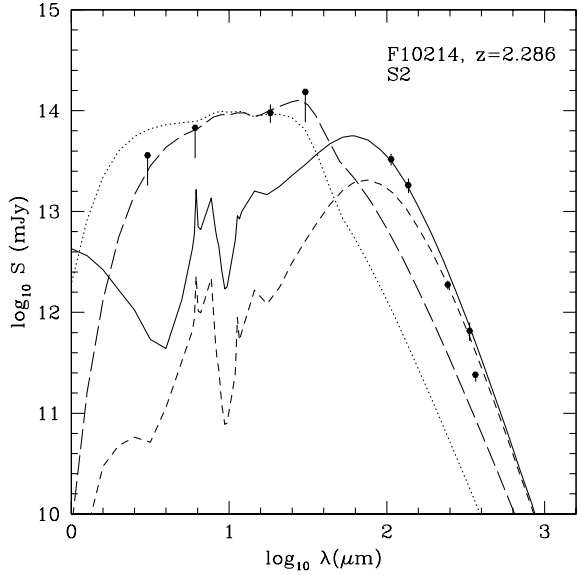,angle=0,width=8cm}
\caption{
Observed (rest-frame) spectral energy distribution for F10214, modelled with M82-type starburst
(solid curve), Arp 220-type starburst (broken curve), AGN dust torus (dotted curve, modified AGN
dust torus model - long-dashed curve).}
\end{figure}

\begin{figure}
\epsfig{file=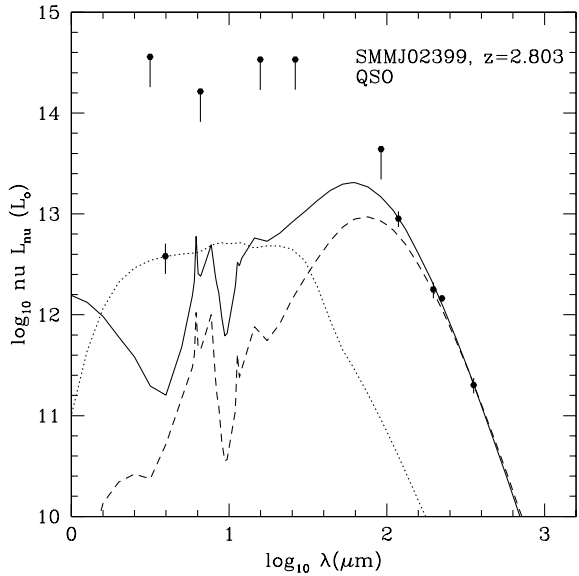,angle=0,width=8cm}
\caption{
Observed spectral energy distribution for SMMJ02399, notation as for Fig 1.}
\end{figure}

\begin{figure}
\epsfig{file=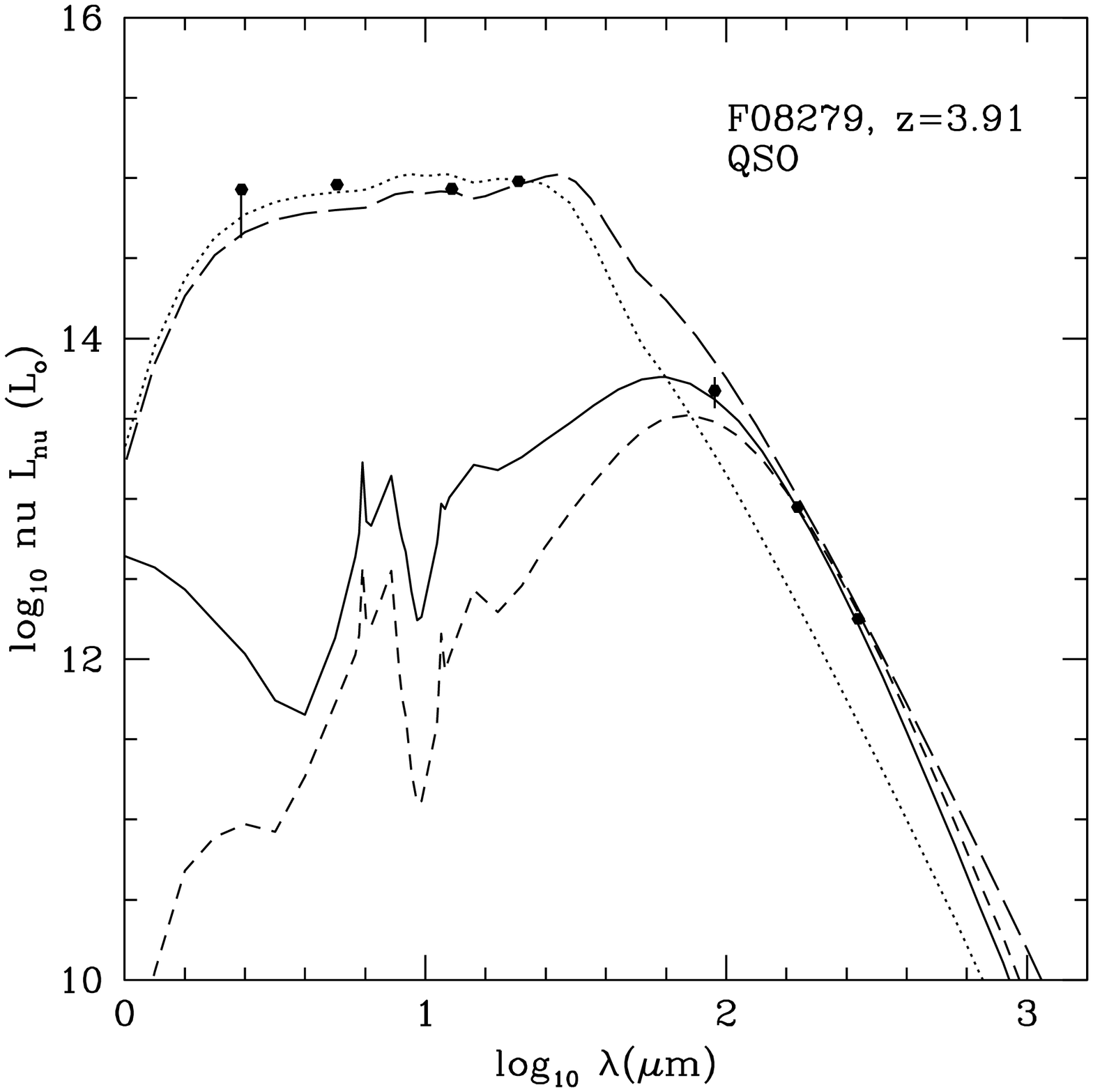,angle=0,width=8cm}
\caption{
Observed spectral energy distribution for F08279, notation as for Fig 1.}
\end{figure}

\begin{figure}
\epsfig{file=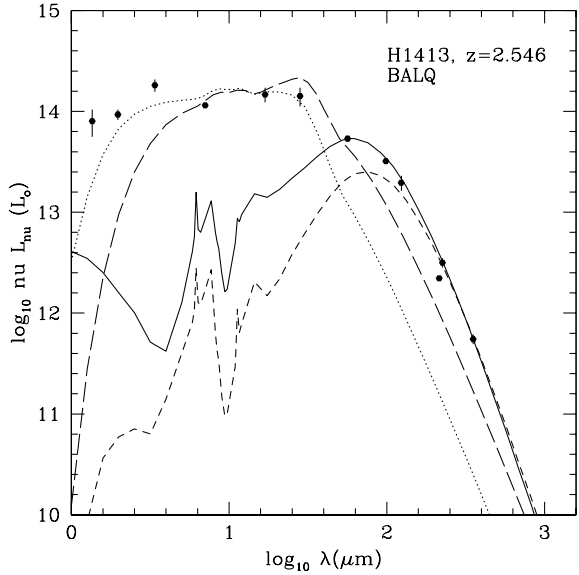,angle=0,width=8cm}
\caption{
Observed spectral energy distribution for H1413, notation as for Fig 1.}
\end{figure}

\begin{figure}
\epsfig{file=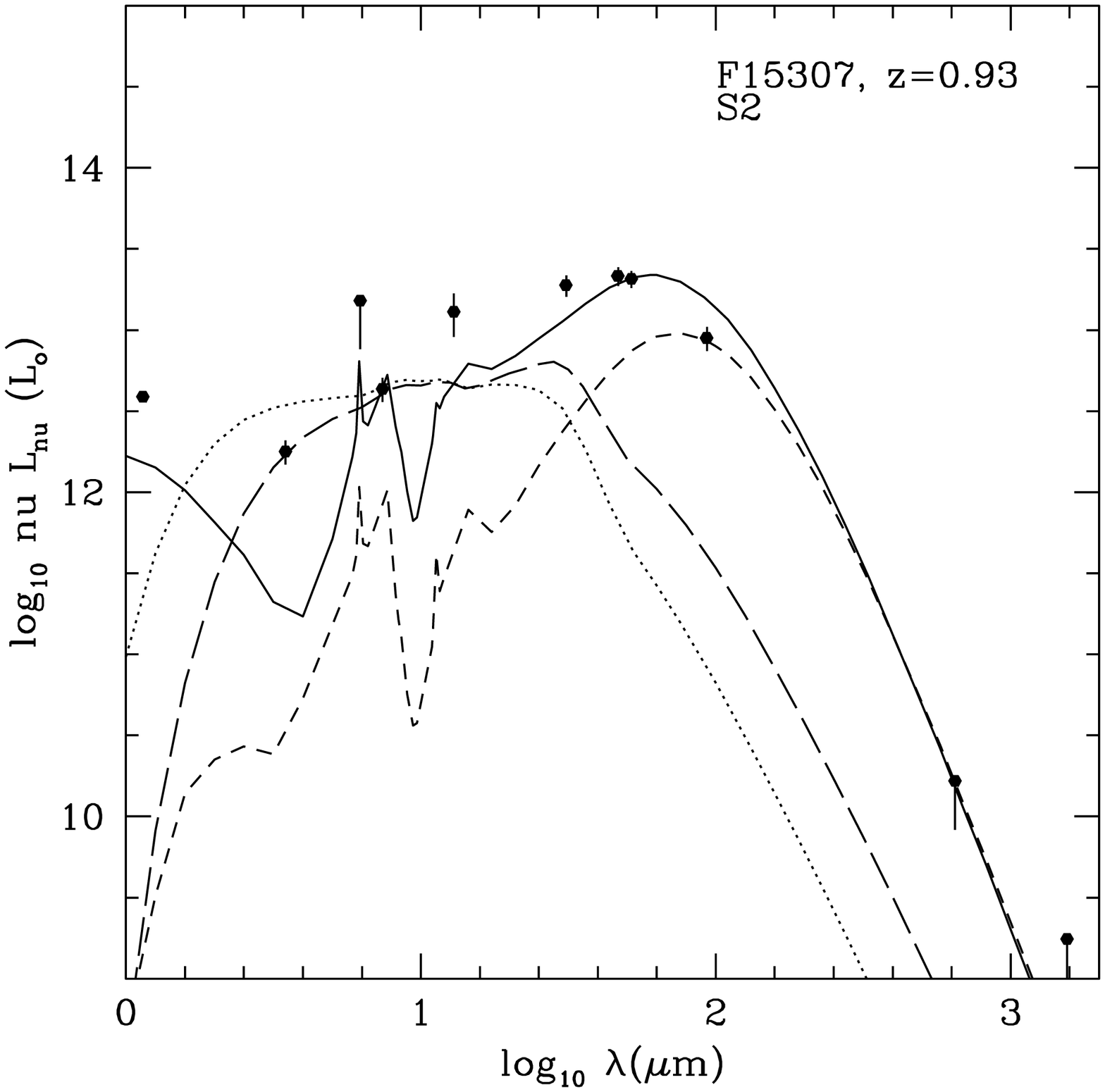,angle=0,width=8cm}
\caption{
Observed spectral energy distribution for F15307, notation as for Fig 1.}
\end{figure}

\begin{figure}
\epsfig{file=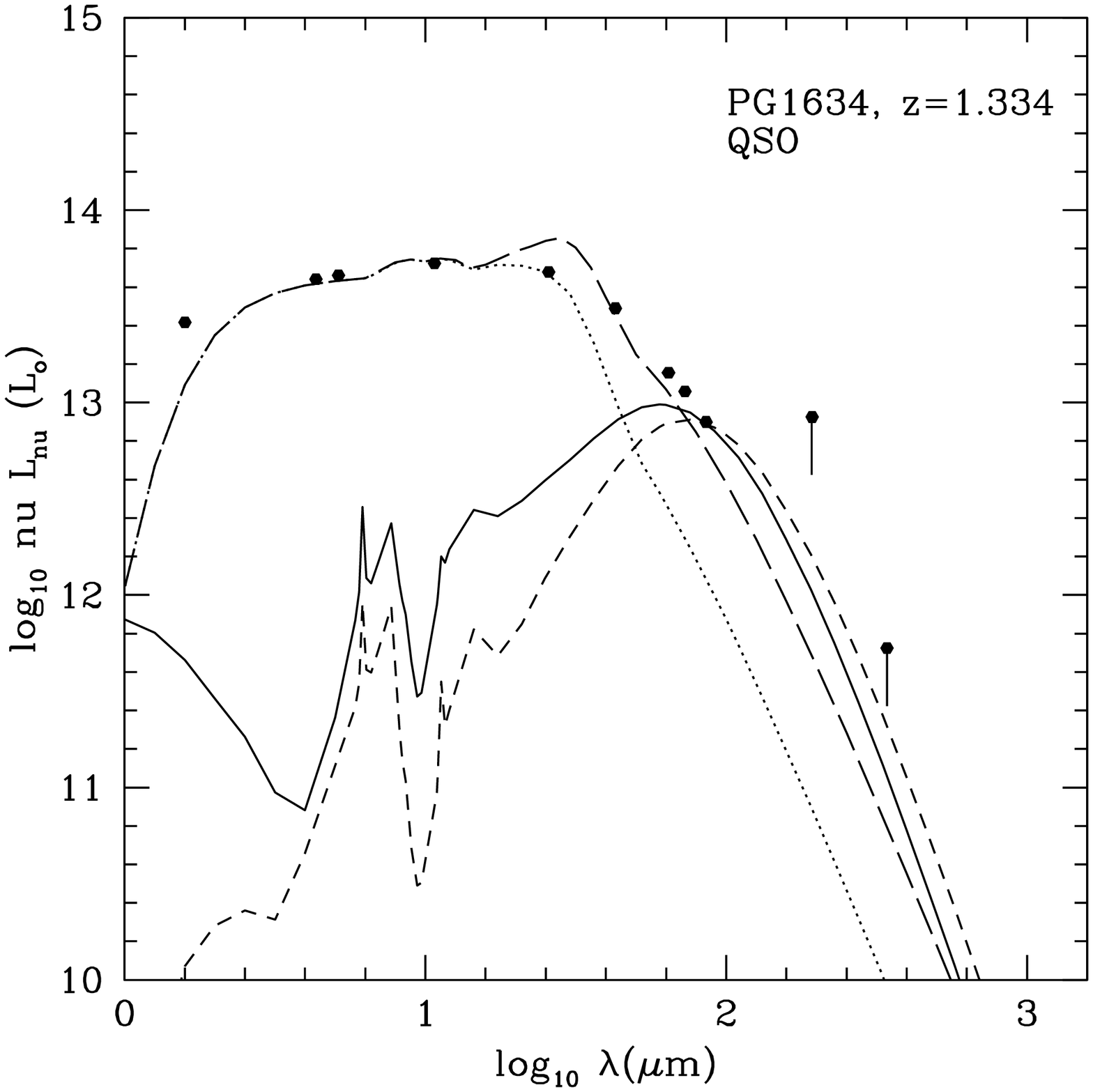,angle=0,width=8cm}
\caption{
Observed spectral energy distribution for PG1634, notation as for Fig 1.}
\end{figure}

Although lensing may affect some of the most luminous objects, there is strong evidence 
for a population of galaxies with far ir luminosities (after correction) in the 
range 1-3x$10^{13} h_{50}^{-2} L_{\odot}$ .   I have argued that in most cases the 
rest-frame radiation longward 
of 50 $\mu$m comes from a starburst component.  The luminosities are such 
as to require star formation rates in the range 3-10x$10^{3} h_{50}^{-2} M_{\odot} \/$yr, 
which would in turn generate most of the heavy elements in a $10^{11} M_{\odot}$ 
galaxy in $10^{7} -10^{8}$ yrs.  Most of 
these galaxies can therefore be considered to be undergoing their most significant 
episode of star formation, ie to be in the process of `formation'.

It appears to be significant that a large fraction of hyperluminous ir galaxies are Seyferts, 
radio-galaxies or QSOs.  This is in part a selection effect, in that luminous AGN have been
selected for submm photometric follow-up.  For the 12 objects found from direct follow-up
of IRAS samples or 850 $\mu$m surveys (Rowan-Robinson 2000a), 50 $\%$ are QSOs or Seyferts and 50 $\%$ 
are narrow-line objects, similar to the proportions found for ultraluminous ir galaxies.
However despite the high proportion
of ultraluminous and hyperluminous galaxies which contain AGN, this does not prove that an
AGN is the source of the rest-frame far infrared radiation.  The ISO-LWS mid-infrared
spectroscopic programme of
Genzel et al (1998) has shown that the far infrared radiation of
most ultraluminous galaxies is powered by a starburst, despite the presence of an AGN
in many cases.  Wilman et al (1999) have shown that the X-ray emission from several 
hyperluminous galaxies is too weak for them to be powered by a typical AGN.

In the Sanders et al (1989) picture, the far infrared and submillimetre emission would 
simply come from the outer 
regions of a warped disk surrounding the AGN.  Some weaknesses of this picture 
as an explanation of the far 
infrared emission from PG quasars have been highlighted by Rowan-Robinson (1995).   
A picture in which both a 
strong starburst and the AGN activity are triggered by the same interaction or merger 
event is far more likely 
to be capable of understanding all phenomena (cf Yamada 1994, Taniguchi et al 1999).

Where hyperluminous galaxies are detected at rest-frame wavelengths in the range 3-30 $\mu$m
(and this can correspond to observed wavelengths up to 150 $\mu$m), the infrared spectrum is often found
to correspond well to emission from a dust torus surrounding an AGN (eg Figs 1, 3, 4, 6).  
This emission often contributes
a substantial fraction of the total infrared (1-1000 $\mu$m) bolometric luminosity.  The
advocacy of this paper for luminous starbursts relate only to the rest-frame emission at 
wavelengths $\geq 50 \mu$m.  Figure 8 shows the correlation between the
luminosity in the starburst component, $L_{sb}$, and the AGN dust torus
component, $L_{tor}$, for hyperluminous infrared galaxies, PG quasars
(Rowan-Robinson 1995), and IRAS galaxies detected in all 4 bands
(Rowan-Robinson and Crawford 1989) (this extends Fig 8 of Rowan-Robinson 1995).  The range of 
the ratio between these
quantities, with 0.1 $\leq L_{sb}/L_{tor} \leq$ 10, is similar over a very wide range of infrared 
luminosity (5 orders
of magnitude), showing that the proposed separation into these
two components for hyperluminous ir galaxies is not at all implausible.

\begin{figure*}
\epsfig{file=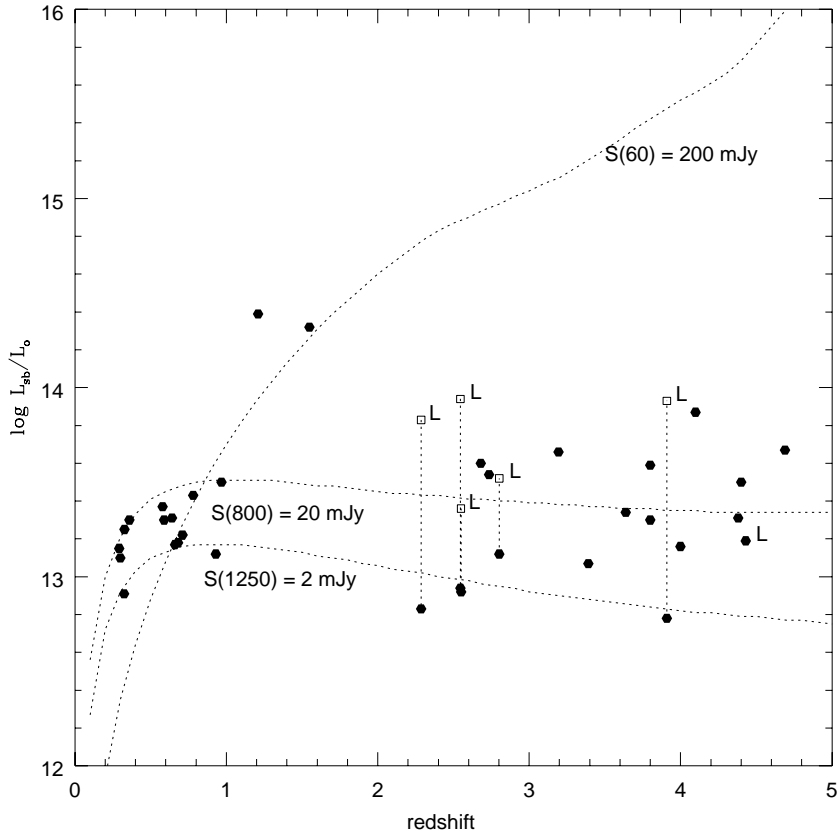,angle=0,width=12cm}
\caption{
Bolometric luminosity in starburst component for galaxies with luminosities $> 10^{13} L_{\odot}$
(Tables 1-4).  
The galaxies labelled L are known to be lensed.  Loci corresponding to the limits
set by S(60) = 200 mJy, S(800) = 20 mJy, and S(1250) = 2 mJy are shown.}
\end{figure*}

\begin{figure*}
\epsfig{file=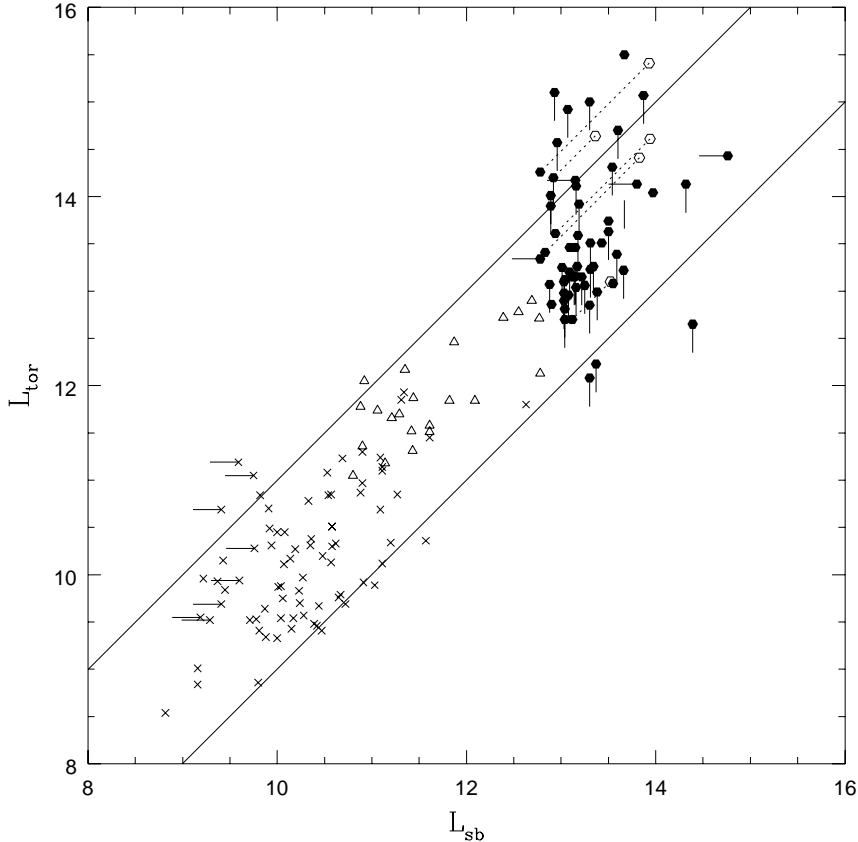,angle=0,width=12cm}
\caption{
Bolometric luminosity in  AGN dust torus component versus bolometric luminosity
in starburst component: filled circles, hyperluminous ir galaxies (this
paper); open triangles, PG quasars (Rowan-Robinson 1995); crosses, IRAS
galaxies detected in 4 bands (Rowan-Robinson and Crawford 1989, galaxies with 
only upper limits on $L_{tor}$ omitted).}
\end{figure*}

The radiative transfer models can be used to derive dust masses and hence, via an
assumed gas-to-dust ratio, gas masses.  For the M82-like starburst model used here the
appropriate conversion is  $M_{dust} = 10^{-4.6} L_{sb}$, in solar units
(Green and Rowan-Robinson 1996).  Rowan-Robinson (2000a) has compared these estimates
with gas mass estimates derived from CO observations, where available, and found good agreement
(within a factor of 2).

The range of ratios of $L_{sb}/M_{gas}$ for hyperluminous galaxies is consistent with that derived for 
ultraluminous starbursts.   
There is a tendency for the time-scale for gas-consumption, assuming a star formation
rate given by eqn (1), to be shorter for the more luminous objects, in the range
$10^7 - 10^8$ yrs (alternatively this could indicate a higher value for the low-mass
cutoff in the IMF).
The cases where a strong limit can be set on $M_{gas}$ are also, generally, those where
the seds do not support the presence of a starburst component.  After correction for the
effects of gravitational lensing, gas masses ranging up to 1-3 x $10^{11} M_{\odot}$
are seen in most hyperluminous galaxies, comparable with the total stellar mass of an $L_*$
galaxy ($10^{11.2} (M/4L) h_{50}^{-2}$).    
Hughes et al (1997) argue that a star-forming
galaxy can not be considered primeval unless it contains a total gas mass of 
$10^{12} M_{\odot}$, but this seems to neglect the fact that 90 $\%$ of the mass of
galaxies resides in the dark (probably non-baryonic) halo.

\section{Conclusions and open issues}

{\bf cirrus}

* PAHs are a powerful diagnostic (uv surface brightness, metallicity)

* need to consider illumination of interstellar dust by old (low-mass) and young (high-mass) stars

* need to model the 3-D dust distribution (cf Silva et al 1998)

* is the cold component far-out dust or giant grains ?

{\bf starbursts}

* PAHS are a powerful diagnostic

* need to model the evolution of a starburst (cf Efstathiou et al 2000)

* need to consider escape of uv radiation durimg the high-$\tau$ phase

* need to consider phase when stars have dispersed the dust

{\bf AGN dust tori}

* lack of 10 $\mu$m feature a powerful constraint (need 10 $\mu$ spectroscopy)

* need proper calculation of cloud model (ie 3-D radiative transfer)

* how far out can dust torus extend ?  $\lambda > 30 \mu$m ?  r $> 350 pc (L_{bol}/10^{13})^{0.5}$ ?

* is $\lambda_{em} \geq 50 \mu$m emission from starburst ?
Sub-mm interferometry could demonstrate the extended nature of the illuminating source.

* there is a need for both an AGN dust torus and starburst components to understand most 
seds of hyperluminous ir galaxies.  
Measured gas masses support, in most cases, the starburst interpretation of rest-frame
far-infrared and submm ($\lambda_{em} \geq 50 \mu$m)emission.

{\bf Arp220-like starbursts}

* what fraction of ULIRGs are high-$\tau$ starbursts ?  This will be important for prospects for
submm sepctroscopy with SCUBA and FIRST.


\begin{thebibliography}{}

\bibitem[1]{} Barvainis R., et al, 1994, Nature 371, 586

\bibitem[2]{} Desert F.-X. et al, 1990, AA 237, 215

\bibitem[3]{} Downes D., et al, 1999, ApJ 513, L1

\bibitem{} Draine B.T. and Lee H.M., 1984, ApJ 285, 89

\bibitem{} Dwek E., 1998, ApJ 501, 643

\bibitem{} Efstathiou A., and Rowan-Robinson M., 1995, MN 273, 649

\bibitem{} Efstathiou A. et al, 2000, MN 313, 734

\bibitem{} Eisenhardt P.R., et al, 1996, ApJ 461,72

\bibitem{} Frayer D.T., et al, 1999, ApJ 514, L13

\bibitem{} Frayer D.T., et al, 1998, ApJ 506, L27

\bibitem{} Genzel R. et al, 1998, 498, 579

\bibitem{} Goldader J.D., Joseph R.D., Doyon R., Sanders D.A., 1997, ApJ 474, 104

\bibitem{} Granato G.L. and Danese L., 1994, MNRAS 268, 235

\bibitem{} Granato G.L., Danese L., Franceschini A., 1997, ApJ 460, L11

\bibitem{} Granato G.L., Danese L., and Franceschini A., 1996, ApJ 460, L11

\bibitem{} Green S. and Rowan-Robinson M., 1996, MNRAS 279, 884

\bibitem{} Hughes D.H., Dunlop J.S., Rawlings S., 1997, MN 289, 766

\bibitem{} Ivison R., et al, 2000, MN (in press), astro-ph/9911069

\bibitem{} Kroker H., et al, 
1996, ApJ 463, L55

\bibitem{} Krugel E. and Siebenmorgen R., 1994, AA 282, 407

\bibitem{} Mathis J.S., Rumpl W., Nordsieck K.H., 1977, ApJ 217, 425

\bibitem{} Pier A. and Krolik J., 1992, ApJ 401, 99

\bibitem{} Rigopoulou D., et al, 1996, MN 278, 1049

\bibitem{} Rowan-Robinson M., 1980, ApJS 44, 403

\bibitem{} Rowan-Robinson M., 1986, MN 219, 737

\bibitem{} Rowan-Robinson M. and Crawford J., 1989, MN 238, 523

\bibitem{} Rowan-Robinson, M., 1992, MN 258, 787

\bibitem{} Rowan-Robinson M. et al, 1993, MN 261, 513

\bibitem{} Rowan-Robinson M., 1995, MN 272, 737

\bibitem{} Rowan-Robinson, M., Efstathiou, A., 1993, MN 263, 675

\bibitem{} Rowan-Robinson M. et al, 1997, MN 289, 490

\bibitem{} Rowan-Robinson M., 2000a, MN 316, 885

\bibitem{} Sanders D.B., et al, 1988, ApJ 325, 74

\bibitem{} Sanders D.B., et al, 1989, ApJ  347, 29

\bibitem{} Scoville, N.Z., Young, J.S., 1983, Ap.J 265, 148

\bibitem{} Siebenmorgen R., Krugel E., 1992, AA 259, 614

\bibitem{} Silva L., Granato G.L., Bressan A., Danese L., 1998, ApJ 509, 103

\bibitem{} Taniguchi Y., Ikeuchi S., Shioya Y., 1999, ApJ 514, L9

\bibitem{} Wilman R.J., et al., 1998, 300, L7

\bibitem{} Yamada T., 1994, ApJ 423, L27

\bibitem{} Xu C. et al, 1998, ApJ 508, 576

\bibitem{} Yorke H.W., 1977, AA 58, 423

\bibitem{} Yun M.S., et al, 1997, ApJ 479, L9

\end{thebibliography}
\end{document}